# Gamma Ray Bursts: Not so Much Deadlier than We Thought

Brian C. Thomas[1, a], Dimitra Atri[2] and Adrian L. Melott[3]

1) Department of Physics and Astronomy, Washburn University, Topeka, KS 66621 USA, brian.thomas@washburn.edu

2) Center for Space Science, New York University Abu Dhabi, PO Box 129188, Abu Dhabi, UAE

3) Department of Physics and Astronomy, University of Kansas, Lawrence, KS 66045 USA

ABSTRACT

We analyze the additional effect on planetary atmospheres of recently detected gamma-ray burst afterglow photons in the range up to 1 TeV. For an Earth-like atmosphere we find that there is a small additional depletion in ozone versus that modeled for only prompt emission. We also find a small enhancement of muon flux at the planet surface. Overall, we conclude that the additional afterglow emission, even with TeV photons, does not result in a significantly larger impact over that found in past studies.

1 INTRODUCTION

   Given the large luminosity of gamma-ray bursts (GRBs) there has been interest for some time in their potential effects on terrestrial-type planets primarily through atmospheric ionization (Thorsett 1995; Scalo & Wheeler 2002; Thomas et al. 2005; Melott & Thomas 2011). High-energy radiation from a variety of sources (GRBs, supernovae, and potentially active galactic nuclei) causes atmospheric ionization and dissociation, which results in reduction of the ozone ($O_3$) layer, exposing near-surface life forms to damaging UVB radiation. As these effects may include mass extinctions (Melott et al. 2004; Melott & Thomas 2009), there has been interest in their relation to the habitability of planets in the Galaxy as well as cosmological constraints on the evolution of life (Martin et al. 2010; Piran & Jiminez 2014; Gowanlock 2016; Piran et al. 2016; Sloan et al. 2017; Melott & Thomas 2018; Lingam et al. 2019; Lingam & Loeb 2019). Until recently, GRBs were known to have radiation up to a few hundred MeV, which can lead to significant ozone depletion at probable distances. An event at 2 kpc from the Earth would have severe effects, possibly catastrophic.

Recent observations have drastically altered the picture of long-burst GRB radiation (MAGIC Collaboration 2019a, 2019b; Abdalla et al. 2019). In the afterglow phase, approximately one day, significant energy is emitted in the form of photons in the range 1 keV to 1 TeV. The most important new component is that in excess of 100 GeV. This renders past modeling of atmospheric effects incomplete.

2 INCLUDING AFTERGLOW EMISSION

In this work we examine the effect of such a GRB in our own galaxy at 2 kpc, the distance found (Thomas et al. 2005) to be the threshold for significant damage from burst photons in past work. Here we add the effects of afterglow photons up to 1 TeV, the range of significant detection. A galactic GRB at 2 kpc is at about $10^{-6}$ less distant than the observed ones; Zyla et al. (2020) and De Angelis et al. (2013) imply that the optical depth of the photons we are considering in the disc of the galaxy is less than one, so it is appropriate to use the spectrum assuming no attenuation (MAGIC Collaboration 2019a). We follow the afterglow development for one day, based on the observations (MAGIC Collaboration 2019a, 2019b; Abdalla et al. 2019). The total energy of the hard afterglow photons is comparable to the energy in the softer X-ray/gamma-ray prompt emission. We combine the prompt and afterglow irradiation in one event, as it is known (Ejzak et al. 2007) that integrated effects work in the long term as a simple sum over inputs. Our computations follow the procedures previously used (Thomas et al. 2005; Ejzak et al. 2007). As in previous modeling, the prompt emission is expected to provide 100 kJ m$^{-2}$ at the Earth, as before using the Band spectrum with a typical peak at 187.5 keV.

The afterglow input is based on observations and modeling reported from recent observations (MAGIC Collaboration 2019b). We take two cases. First, we use the spectrum for photons of energy 1 keV $\leq$ E $\leq$ 1 TeV from MAGIC Collaboration (2019b) between 68 and 180 seconds, with a total energy of 4 x 10$^{45}$ J. We then take a second case in which the afterglow is extrapolated to 1 day, with total energy 8 x 10$^{45}$ J, based on the same observations and a review of previous work. These two cases taken at 2 kpc then add to the prompt fluence about 80 kJ m$^{-2}$ and 160 kJ m$^{-2}$, respectively. However, it is important to note that this fluence comes in much higher energy photons compared to the prompt emission.

3 ATMOSPHERIC IONIZATION AND CHEMISTRY MODELING

Ionization rate profiles are calculated separately from the atmospheric model, following the method used in Gehrels et al. (2003), Thomas et al. (2005) and Ejzak et al. (2007). The total photon flux in each of 88 energy bins in the range 10$^{-3}$ MeV $\leq$ E $\leq$ 10$^6$ MeV is propagated vertically through a standard atmosphere (adjusted for the appropriate latitude and time when input to the atmospheric model), attenuated with altitude by an exponential decay law with energy-dependent absorption coefficients taken from a lookup table. The lookup table values for 10$^{-3}$ MeV $\leq$ E $\leq$ 10$^5$ MeV were obtained from the National Institute of Standards and Technology (NIST) XCOM database, available online (Berger et al. 2005), based on a mixture of 79% N$_2$ and 21% O$_2$. That database does not extend above 10$^5$ MeV, so we have generated values 10$^5$ MeV < E $\leq$ 10$^6$ MeV using a log-log extrapolation based on the database values between 10$^4$ MeV and 10$^5$ MeV. The energy deposited in each atmospheric layer is computed and then converted to an ionization rate using 35 eV per ion pair (Porter et al. 1976); here "ion pair" refers to a variety of products involving excited N-atom states plus positively charged N ions, all of which collectively contribute to the subsequent chemistry that results in depletion of $O_3$. The vertical ionization rate profiles are then mapped onto the altitude and latitude grid used by the GSFC model. More details can be found in Thomas et al. (2005).

In this work we take a GRB occurring over Earth's equator, in late June (around the Northern Summer solstice). Both the latitude over which the GRB occurs and the time of year has an effect on geographic distribution and overall magnitude of $O_3$ depletion. For an equatorial burst the effect is roughly symmetric around the equator, rather than being concentrated in a given hemisphere as is true for a burst occurring over a pole. The time of year affects distribution and magnitude of depletion through photochemical reactions in the polar regions. An event in June tends to lead to larger overall depletion. Therefore, the case we have chosen represents a fairly uniform depletion in latitude, with a total depletion value toward the upper end of the likely range. More detailed discussion of these factors can be found in Thomas et al. (2005).

Atmospheric chemistry modeling was performed using the Goddard Space Flight Center (GSFC) 2D (latitude-altitude) chemistry and dynamics model. This model has been extensively tested for cases similar to the work presented here (Thomas et al. 2005, 2007, 2008; Ejzak et al. 2007). The model runs from the ground to 116 km in altitude, with approximately 2 km altitude bins, and from pole-to-pole in 18 bands of 10-degree latitude each. The model includes 65 chemical species, 37 transported species and "families" (e.g. $NO_y$), winds, small scale mixing, solar cycle variations, and heterogeneous processes (including surface chemistry on polar stratospheric clouds and sulfate aerosols). We use the model in a pre-industrial state, with anthropogenic compounds (such as CFCs) set to zero.

Ionization profiles generated as described above are read into the GSFC model as production sources of $NO_y$ and $HO_x$. It is assumed that for each ion-electron pair produced, 1.25 $NO_y$ molecules are produced at all pressure levels (Porter et al. 1976) and 2.0 molecules of $HO_x$ are produced below 75 km and less than 2.0 (from 1.99 to 0.0) for altitudes greater than 75 km (Solomon et al. 1981). Ionization is input for a single, one day timestep. The model is then run for 20 years, long enough for the atmosphere to recover back to pre-burst equilibrium.

## 3.1 ATMOSPHERIC MODELING RESULTS

The simplest way to compare $O_3$ depletion between cases is to look at the globally averaged change. Figure 1 shows this for the prompt-only case and for both prompt+afterglow cases. Maximum depletion is reached about 2 years after the burst due to effects of transport and seasonality (Thomas et al. 2005). This maximum depletion value ranges from about 40% in the prompt-only case to about 44% in the case with the higher fluence afterglow. Figure 2 shows the geographical (latitude) distribution of depletion over time.

The relatively small increase in $O_3$ depletion may be surprising given that the total fluence is more than doubled over the prompt-only case. However, this can be explained as follows. One might expect higher energy photons to penetrate more deeply in the atmosphere and therefore have a larger effect. This is true to a point, but the photon attenuation depth in air peaks around photon energies of 40-50 MeV and then tapers off (Zyla et al. 2020). Therefore, adding photons above this energy actually results in energy deposition at higher altitudes, which has some effect on $O_3$, but not as much as might be expected. Figure 3 shows the ionization profiles for all three cases. Notice that the afterglow cases show more ionization overall, but the maximum ionization region shifts to higher altitude, thereby limiting the additional impact. In addition, there is an asymptotic effect for depletion and the relation between globally averaged depletion and total fluence (for a fixed spectrum) is roughly cubic, not linear (Thomas et al. 2005); a similar relationship has been noted for solar energetic particles (Lingam & Loeb 2017).

## 4 MODELING MUON SURFACE EXPOSURE

High-energy gamma-rays can produce secondary particles at a planet's surface due to interactions with the atmosphere. Muons are the most biologically important product. Atri et al. (2014) examined this effect for GRB prompt emission and found a negligible effect. Here we modeled the interaction of the GRB afterglow photons with the atmosphere using the CORSIKA package, a widely used code in the astroparticle physics community (Heck et al., 1998). It is a Monte Carlo code that simulates the propagation of high-energy charged particles and photons with the atmosphere. The code is regularly calibrated with latest experimental results, which makes it ideal for our calculations. We have used CORSIKA for similar calculations earlier, where we calculated gamma ray-induced muon flux for energies up to 10 GeV (Atri et al, 2014). We extend those results here by following the same method for photons of energies up to 1 TeV.

4.1 MUON MODELING RESULTS

At each primary energy simulations with $10^9$ photons were carried out and the average number of muons produced at the ground level calculated for that primary energy ($10^9$ photons were used for simulations to minimize statistical error and also ensure that simulations finish in a reasonable timeframe). Those results were then combined with the photon spectrum described above to yield the average number of muons in the shower reaching the ground level, shown in Figure 4 as a function of the incident photon energy.

Unlike charged particles, gamma rays are very inefficient at producing muons (Atri et al., 2011). As it can be seen in Figure 4, the flux of muons at the ground level is several orders of magnitude smaller compared to the incident photon flux at low energies, and about an order of magnitude lower at 1 TeV. Overall, we obtained a total of $1.59 \times 10^{-11}$ muons $cm^{-2} s^{-1}$ on the surface. This is because the flux of high-energy gamma rays is extremely small. The total flux is a billion times smaller compared to the background muon flux, which is about $1.5 \times 10^{-2}$ muons $cm^{-2} s^{-1}$. We therefore conclude that GRB-induced muons do not have any biological impact.

5 CONCLUSIONS

In light of new observations of very high energy photon emission from GRBs we revisited the question of the potential impact of a relatively nearby GRB on life on Earth, or other terrestrial-type planets with an Earth-like atmosphere. Despite an increase in the total energy fluence when including afterglow emission, our modeling shows only a modest increase in ozone depletion. In addition, the flux of secondary muons at ground-level is found to be too small to have an impact on life. We conclude that, in general, the discovery of TeV afterglow emission does not significantly increase the threat from nearby GRBs on life on Earth or other terrestrial-type planets.

ACKNOWLEDGEMENTS

We thank G.W. Wilson and R.J. Scherrer for helpful inputs in the preparation for this work and writing of the manuscript. DA acknowledges support from the New York University Abu Dhabi (NYUAD) Institute research grant G1502. This research was carried out on the High Performance Computing resources at NYU Abu Dhabi. Analysis and plotting was done using the NCAR Command Language (Version 6.6.2, Boulder, Colorado: UCAR/NCAR/CISL/TDD. http://dx.doi.org/10.5065/D6WD3XH5).

DATA AVAILABILITY

*The data underlying this article are freely available at https://zenodo.org/record/4073593*

REFERENCES

Abdalla H. et al., 2019, Nature, 575, 464

Atri D., Melott A.L., Karam, A, 2014, International Journal of Astrobiology, 13.3, 224

Atri D. & Melott A.L., 2011, Radiation Physics and Chemistry, 80.6, 701

Berger M.J. et al., 2005, XCOM: Photon Cross Section Database, Version 1.3 (Gaithersburg: NIST), http://physics.nist.gov/xcom

De Angelis A., Galanti G., Roncadelli M., 2013, MNRAS, 432, 3245

Ejzak L.M, Melott A.L., Medvedev M.V., Thomas B.C., 2007, ApJ, 644, 373

Gowanlock M.G., 2016, ApJ, 832, 38

Heck D. et al., 1998, Report fzka, 6019.11

Lingam M. & Loeb A., 2017, ApJ, 848, 41

Lingam M. & Loeb A., 2019, Reviews of Modern Physics, 91, 021002

Lingam M., Ginsburg I., Bialy S., 2019, ApJ, 877, 62

MAGIC Collaboration, 2019a, Nature, 575, 455

MAGIC Collaboration, 2019b, Nature, 575, 459

Martin O., Cardenas R., Guimarais M., Penate L., Horvath J., Galante D., 2010, Astrophys. Space Sci., 326, 61

Melott A.L., Lieberman B.S., Martin L.D., Medvedev M.V., Thomas B.C., Cannizzo J.K., Gehrels N., Jackman C.H., 2004, Int. J. Astrobiol., 3, 55

Melott A.L. & Thomas B.C., 2009, Paleobiology, 35, 311

Melott A.L. & Thomas B.C., 2011, Astrobiology, 11, 343

Melott A.L. & Thomas B.C., 2018, Lethaia, 51, 325

Piran T. & Jiminez R., 2014, PRL, 113, 231102

Piran T., Jiminez R., Cuesta A.J., Simpson F., Verde L., 2016, PRL, 116, 081301

Scalo J. & Wheeler J.C., 2002, ApJ, 566, 723

Sloan D., Batista R.A., Loeb A., 2017, Scientific Reports, 7, 5419

Thomas B.C. et al., 2005, ApJ, 634, 509

Thorsett S.E., 1995, ApJL, 444, L53

Zyla P.A. et al. (Particle Data Group), 2020, to be published in Prog. Theor. Exp. Phys. 2020, 083C01

FIGURES

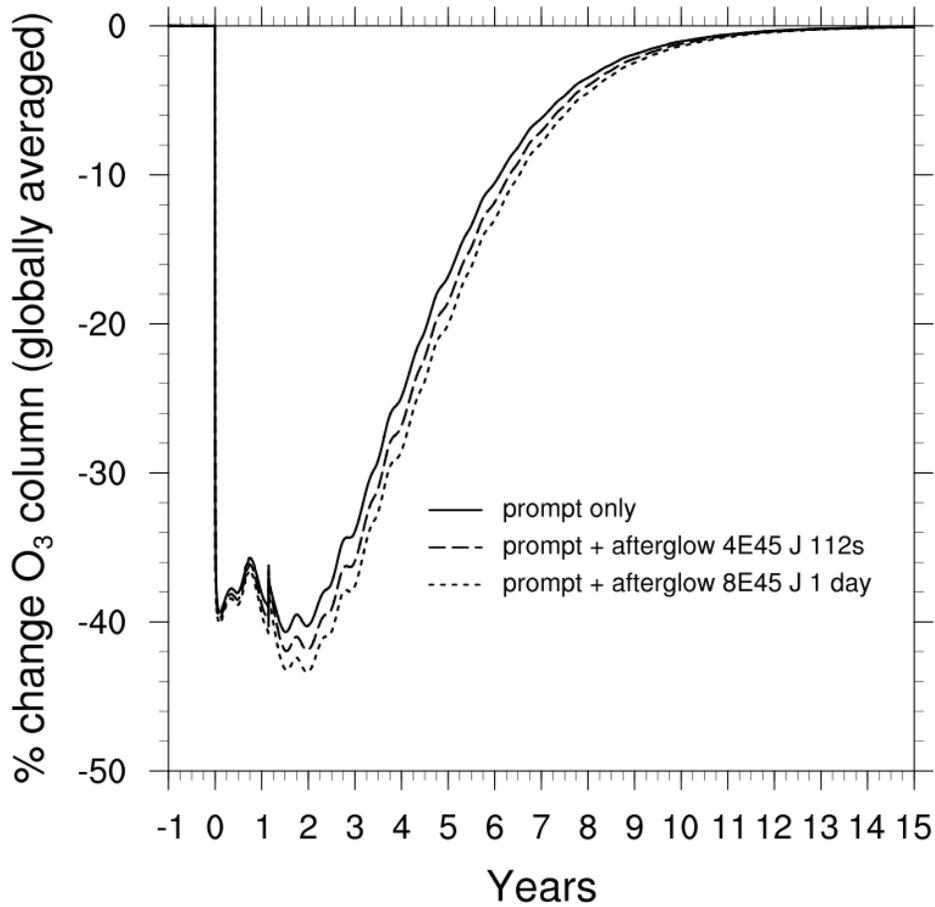

Figure 1 – Globally averaged percent difference (case vs control) of $O_3$ column density, for each case described Section 2, starting 1 year before the burst.

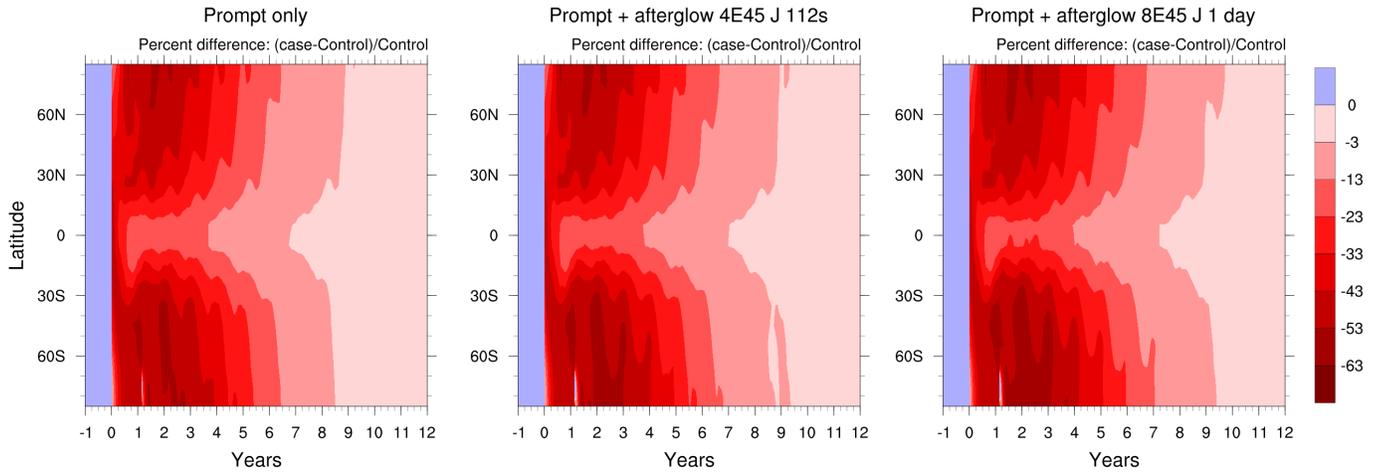

Figure 2 – Percent difference (case vs control) of $O_3$ column density as a function of latitude and time, for each case described in Section 2, starting 1 year before the burst.

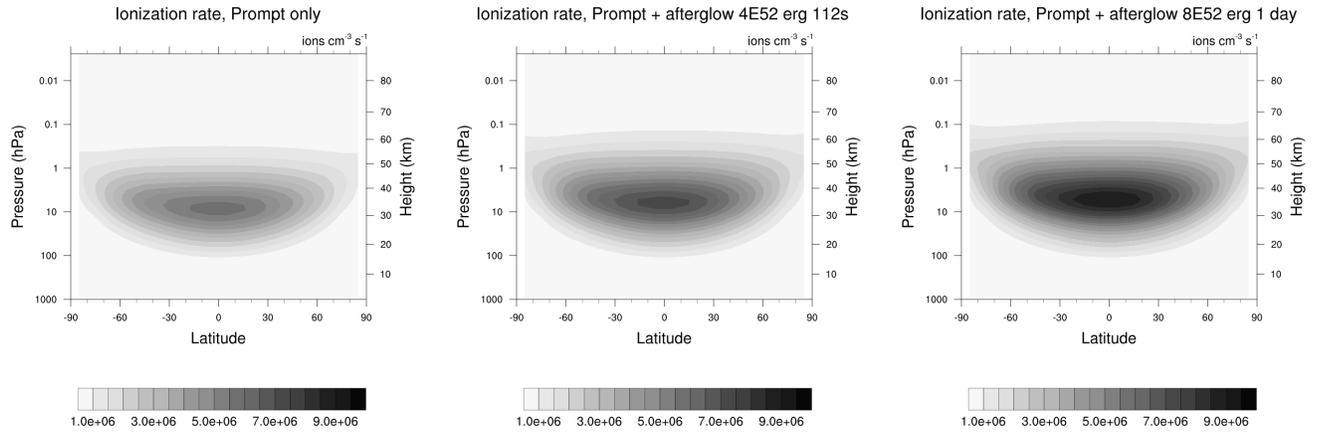

Figure 3 – Ionization rate profiles (as a function of altitude and latitude) for each case described in Section 2

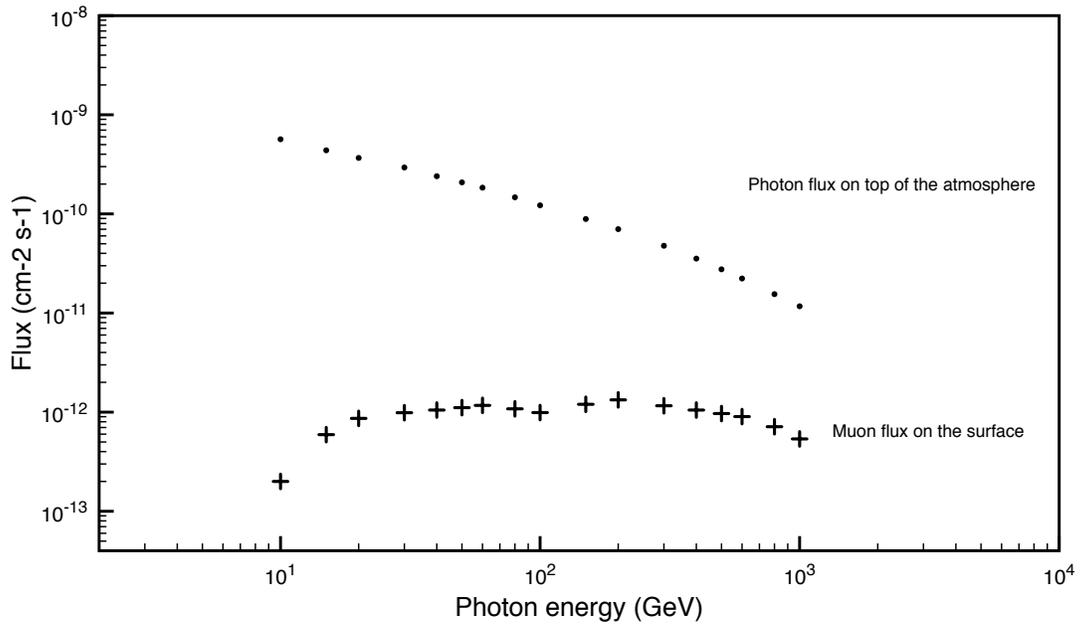

Figure 4 – Flux of muons at ground level as a function of afterglow incident photon energy.